# Organizational Learning in Industry 4.0: Applying Crossan's 4I Framework with Double Loop Learning

*Short Paper*

**Nimra Akram**
The University of Melbourne
Australia
nakram@student.unimelb.edu.au

**Atif Ahmad**
The University of Melbourne
Australia
atif@unimelb.edu.au

**Sean B Maynard**
The University of Melbourne
Australia
seanbm@unimelb.edu.au

## Abstract

*The Advanced Dynamic Security Learning (DSL) Process Model is an Industry 4.0 cybersecurity incident response architecture proposed in this paper. This model addresses proactive and reflective cybersecurity governance across complex cyber-physical systems by combining Argyris and Schön's double-loop learning theory with Crossan's 4I organizational learning framework. Given that 65% of industrial companies suffer ransomware attacks annually and many of them lack cybersecurity awareness, this reveals the gravity of cyber threats. Feedforward and feedback learning loops in this paradigm help promote strategic transformation and ongoing growth. The DSL model helps Industry 4.0 organizations adapt to growing challenges posed by the projected 18.8 billion IoT devices by bridging operational obstacles and promoting systemic resilience. This research presents a scalable, methodical cybersecurity maturity approach based on a comprehensive analysis of the literature and a qualitative study.*

## Introduction

Industry 4.0 has transformed sectoral operations by combining automation, cloud computing, big data, cyber-physical systems, and the Internet of Things. These technologies together form an ecosystem that allows resource management, predictive maintenance, and real-time decision-making. Global manufacturers have initiated Industry 4.0 strategies, and the number of connected IoT devices is projected to reach 18.8 billion (Sinha, 2024). Recent high-profile attacks like Colonial Pipeline and Maersk demonstrate the compromise of both operational systems and broader infrastructures by cyberattacks in Industry 4.0. With 65% of manufacturing organizations suffering ransomware attacks in the past year, integrating cross-domain coordination strategies is critical to effective incident response (Mahendru, 2024). Division of incident response duties requires an integrated learning model to provide alignment, consistency, and shared knowledge across all organizational levels (Ahmad et al., 2021). Current models tend to focus on IT-centric environments with static digital assets and overlook the dynamic, decentralized nature of industrial operational technology (OT) systems. This creates a critical gap in how organizations prepare for and respond to threats in cyber-physical environments. Many models applied in industries such as banking and healthcare are too generic or only relevant to sectors with concentrated digital assets and limited cyber-physical connections.

This paper addresses this gap by proposing an Advanced Dynamic Security Learning (DSL) model, a novel incident response model that integrates Crossan's 4I framework with Argyris and Schön's double-loop learning to build adaptive, learning-driven cybersecurity resilience in Industry 4.0. The proposed Advanced DSL model extends Ahmad et al.'s (2015) DSL model as it moves beyond the single-domain approach to dual-domain learning architecture to address Industry 4.0's unique cyber-physical challenges and geographically distributed operations. The research question is "How can organizations leverage the 4I framework along with double-loop learning to transform their cybersecurity incident response capabilities in the Industry 4.0 environment?". It emphasizes cyber resilience and organizational learning in industry environments growing increasingly digital with an emphasis on systemic transformation and adaptive capabilities. This research aims to provide a comprehensive, context-specific model that combines double-loop learning and the 4I framework into incident response techniques helping industries in facing and reacting to hazards. The research will also look at how knowledge of incident response at the plant level may be shared and institutionalized at the corporate level, therefore linking localized action and enterprise-wide policy reform. The structure of the following chapters starts with literature review of organizational learning theories and cybersecurity incident response methods, systematic thematic analysis methodology, the proposed Advanced DSL model, the potential research contribution, limitation and future research direction. This study's novel contribution in comparison to existing cybersecurity learning models includes: (1) dual-domain SOC architecture like IT-OT convergence, (2) real-time feed-forward and feedback integration and (3) cross-stakeholder knowledge synthesis.

# Literature Review

## *Cybersecurity Challenges and Trends in Industry 4.0*

Lezzi et al. (2018) argue that the shift from isolated systems to highly linked architectures especially when OT systems, which are meant for dependability and longevity; are exposed to external networks without enough protections makes conventional cybersecurity techniques insufficient. Due to legacy OT systems' longer operational lives and limited security patches, they are particularly susceptible. Outdated software, open routes of communication, and poor IT-OT network segmentation could all jeopardize critical infrastructure. Recent research indicates a 400% increase in malware attacks targeting IoT devices in 2023, underscoring the need for faster detection in industrial environments (McKendrick, 2023). Approximately 26% of companies provide no cybersecurity training for their employees. (Hornetsecurity, 2024). These numbers show how lack of cybersecurity knowledge puts industrial ecosystems at risk for operational, financial, and reputational harm. This persistent misalignment between technological risk and organizational response exposes a key limitation in traditional learning approaches. They fail to integrate adaptive, cross-functional learning mechanisms that can evolve with the threat landscape. This difference between perceived importance and actual preparation affects techniques for incident response. Organizations that want to keep up with Industry 4.0's fast pace need proactive, reflective learning strategies.

## *Incident Response in Cybersecurity*

Incident response (IR) is commonly understood as a sequential six-stage process encompassing: Preparation (establishing policies, training, and tools), Identification (recognizing potential security incidents), Containment (limiting the spread and impact), Eradication (removing the cause of the incident), Recovery (restoring systems and operations), and Follow-up (reviewing and improving future responses). Although NIST SP 800-61 emphasizes preparation, detection, response, and post-incident learning, mentioned (Ahmad et al., 2020), ISO/IEC 27035 promotes organizational readiness and clearly defined responsibilities during an event. Many compliance-driven strategies give checklists and reporting top priority over contextual responsiveness (O'Neill et al., 2021). Industry 4.0's speed and interconnectedness put restrictions on conventional incident response models. Attackers using cloud, IoT, and cyber-physical interface vulnerabilities are very adaptable, therefore incident response plans must alter as well. One problem is the lack of dynamic learning. Incident response systems often ignore policy assumptions, risk tolerances, and communication failures, as they are disconnected from corporate strategy Patterson et al. (2023). Real-time incident threat identification is difficult, and the volume and speed of data produced by smart factories surpass those of conventional monitoring systems, which delays the danger detection and reaction (Lezzi et al., 2018). Many OT and IoT parts depend on outdated, insecure systems that are neither upgradeable nor fixable (Sotolani & Galegale, 2023). Another issue is the lack of cybersecurity experts versed in IT and OT. Companies keep making errors due to the lack of effective shared learning and collaborative practices, therefore impeding the industry's development. IT-OT integration in incident response is crucial but currently in its infancy in Industry 4.0. Although OT settings give safety, dependability, and uptime top priority, typical IT departments focus more on data integrity, confidentiality, and regulatory compliance. Sometimes the cultural gaps between IT and OT teams lead to different incident response protocols, duplicate effort, and delayed crisis communication (Chen, 2013; Ahmad et al., 2021). Unified, adaptable, context-aware, and mutually understood incident response models are needed to surmount these obstacles. These models must enable cooperative reflection and the dissemination of incident information by means of organizational learning processes spanning across departmental lines (Ahmad et al., 2020)

*Organizational Learning*

This section covers organizational learning models, which are the single-loop and double-loop models that provide greater introspection and transformation. A comparative analysis highlights the application of these models in dynamic cyber-physical environments. Additionally covered in this section are the 4I framework for embedding insights across organizational levels. Single loop learning involves seeing and fixing mistakes without changing the values, practices, or laws that control behavior or challenging the beliefs. It emphasizes correcting the actions rather than questioning the underlying assumptions. Single-loop learning, in cybersecurity incident response, is the reactive approach of correcting firewall rules, software vulnerabilities, and security breaches without investigating organizational issues (Shedden et al., 2010). Double-loop learning, developed by Argyris and Schön, questions and may alters the underlying assumptions, rules, and guidelines that resulted in a mistake (Argyris, 1977). Double-loop learning provides the flexibility dynamic threat environments need. The encouragement of research and strategic reflection, organizations may foresee and change their policies and technologies, therefore strengthening resilience. Double-loop learning fits the systematic learning of Industry 4.0 and continuous industry change. Particularly in Industry 4.0, the following **Table 1** compares cybersecurity incident response learning approaches to help one better appreciate their advantages:

| Learning Loop | Focus | Action Taken | Applicable Systems | Learning Depth | Example in Incident Response |
|---|---|---|---|---|---|
| **Single Loop** | Error correction | Fix the issue | Simple and isolated systems (e.g., legacy OT or sensor systems) | Low | Applying security patches to PLC firmware after detecting known vulnerability |
| **Double Loop** | Questioning assumptions | Change policies, procedures or models | Interconnected cyber-physical systems (e.g., SCADA-ERP integration) | Medium to High | Revising remote access policies after identifying failure in VPN segmentation during ransomware event |
| **Table 1. Learning Loops in Cybersecurity IR with Examples of Industry 4.0** | | | | | |

The 4I Framework provides a structured approach to organizational learning through four interconnected processes: Intuiting (Individual level): Recognition of patterns or insights based on experience. Interpreting (Individual/Group level): Explaining insights to others and developing shared understanding. Integrating (Group/Organizational level): Developing coordinated actions through shared mental models. Institutionalizing (Organizational level): Embedding learning into systems, routines, and structures.

This 4I framework captures multilayer learning and is ideal for Industry 4.0 post-incident evaluations. During a cybersecurity attack, front-line engineers could notice odd system activity and bring it up with the team. Integration drives coordination. These lessons are institutionalized when they lead to new policies, training programs, or monitoring tools. According to the 4I framework as mentioned by Crossan et al., (1999) learning must go beyond levels The 4I framework makes both double loops learning feasible as well as the upward and downward flow of information that transforms individual insights into organizational intelligence.

### *Integration of the 4I Framework and Double-Loop Learning in Industry 4.0 Cybersecurity Incident Response*

In the 4I framework, organizational learning is said to involve intuiting, interpreting, integrating, and institutionalizing at the individual, group, and organizational levels (Crossan et al., 1999). There is an inconsistency in the literature where knowledge comes before transformation. Companies should look at why security plans failed, which assumptions were false, and how systematic vulnerabilities were neglected instead of just documenting what happened (Shedden et al., 2010; Ahmad et al., 2020). Salimath and Philip (2020) claim that organizational learning in security management improves long-term resilience and moves the emphasis from containment to transformation. The 4I framework emphasizes that learning must move beyond individual intuition and interpretation toward integration at group level and institutionalization at the organizational level (Crossan et al., 1999). Double-loop learning-based assessments look at reaction priorities, decision-making procedures, and crisis response cultural norms. Reviewing assumptions and reengineering rules helps to lower systematic risk and inspire ongoing improvement (Patterson et al., 2023; Ahmad, 2021). Crossan et al. (1999) argue that maintaining a balance between feed-forward learning (intuition to institutionalization) and feedback learning (institutionalization to intuition) enables strategic renewal and resilience. Domain-wide integration is essential to fulfill the cyber-physical convergence of Industry 4.0 between IT and OT operations; double-loop learning helps to provide communication and reaction systems.

The theoretical advancement of the proposed Advanced DSL model is clearly demonstrated using the **Table 2** which provides the comparative analysis of existing incident response frameworks across five critical dimensions.

| Framework | Conceptualization & Learning Approach | Stakeholders' practices | Decision Making Approach | Lens/Perspective |
|---|---|---|---|---|
| **Traditional NIST SP 800-61 / ISO 27035** | Single loop learning focused on reactive incident response | IR Teams<br>Siloed approach | Checklist-based<br>Compliance-focused | IT-focused with limited operational technology (OT) consideration |
| **Ahmad et al. DSL (2015)** | Single and double-loop learning enabling both corrective actions and underlying assumption questioning | IRT, Security managers, Senior management<br>Cross-functional collaboration<br>Learning-driven practices | Collaborative<br>Security insight-driven<br>Multi-stakeholder negotiation | Organizational Information Systems |
| **Advanced DSL (Proposed)** | Single and double loop learning with enhanced adaptive mechanisms for complex cyber-physical environments | Dual SOC teams (IT/OT)<br>Cross-domain stakeholder panels | Cross-domain coordination<br>Real-time responses | Cyber-physical systems lens addressing convergence of digital and physical domains |

| | | Coordinated practices | Distributed yet coordinated | |
|---|---|---|---|---|
| **Table 2. Analysis of incident response frameworks.** | | | | |

## Methodology

This paper uses a qualitative and exploratory methodology applying Crossan's 4I framework to Industry 4.0 cybersecurity incident response. The central goal was to conceptualize a learning-centric incident response model that addresses the unique operational structure of cyber-physical systems in Industry 4.0 environments. This conceptual development employs literature narrative synthesis of existing literature to generate new theoretical insights. The methodology is based on the structured literature review on the 4I framework, cybersecurity incident response, and organizational learning. Literature selection was guided by a structured keyword strategy: ("incident response" OR "cybersecurity management") AND ("double loop learning" OR "organizational learning" OR "4I framework") AND ("Industry 4.0" OR "operational technology") to identify relevant scholarly and industry publications. These keywords yielded 227 initial results from Google Scholar; 204 articles remained after applying the temporal filter (2015-2025) which perfectly aligns with Industry 4.0's implementation period and the evolution of cyber-physical security threats. The articles were further refined to 50 peer-reviewed articles related to operational technology (OT) and organizational learning. After the selection of literature, this research incorporates insights from published industry reports and case studies that document the cybersecurity incidents in manufacturing, energy, and telecommunication sectors. Published materials analyzed include post-incident audit reports, case studies of major cyber incidents, and academic research addressing organizational learning and threat mitigation. These documents directly informed both the identification of system-level gaps and the structuring of learning loops in the proposed model. To conduct a narrative synthesis, a preliminary synthesis of study findings was developed to explore the relationships within and between the studies. This synthesis assesses how different theoretical frameworks address cybersecurity challenges in smart factories. Three core areas of synthesis were identified: (1) IT-OT learning integration gaps, (2) cross-domain knowledge transfer barriers, and (3) adaptive response limitations in cyber-physical environments. The synthesis process formed the foundation of the Advanced DSL model by integrating these findings into the core category of "cyber-physical organizational learning". The result is the application of 4I framework (Crossan et al., 1999) in the context of Industry 4.0 with additional stakeholder such as junior SOC (IoT) along with double-loop learning in the Advanced DSL model. The Advanced DSL model development follows established Design Science Research principles (Peffers et al., 2007), specifically focusing on the design and development phase where the 4I organizational learning framework is enhanced with double loop learning concepts to address the multi-stakeholder nature of OT incident response. The model employs a Role-Process Learning Matrix structure that systematically maps stakeholder roles against sequential learning processes, facilitating knowledge flow from individual recognition through organizational institutionalization across both IT and OT domains. The conceptual development of process model is the initial stage of planned multi-phase research. To refine, validate and extend the Advanced DSL model, the researcher will collect empirical data through multiple case studies and focus group meetings. Before field testing in the real-world environment of Industry 4.0, this theoretical model gives the strong foundation.

## Proposed Model

This study extends the DSL model proposed by Ahmad et al. (2015) to support the structural and functional changes that the companies operating with Industry 4.0 technologies require. Among them are geographically scattered production facilities, separate operational technology (OT) and information technology (IT) systems, junior Security Operations Center (SOC) teams in both the plant and the head office. A survey of 124 IT and IS professionals from Czech and Slovak companies confirmed that many organizations face challenges managing geographically scattered OT and IT systems effectively (Petrová et al., 2024). The model enables strategic reflection and forward development by linking each phase with organizational actors and learning, either in a single or double-loop. It also follows the reasoning put forward by Crossan et al. (1999), which calls for feed-forward (exploration) and feedback (exploitation).

### *The Advanced DSL Process Model*

From junior SOC teams to executive leadership, the structure shows the 4I learning process through intuiting, interpreting, integrating, and institutionalizing across vertical levels and horizontal organizational functions. Each process examines how the organizations generate, absorbs, shares, combines, and changes learning, illustrated in **Figure 1**:

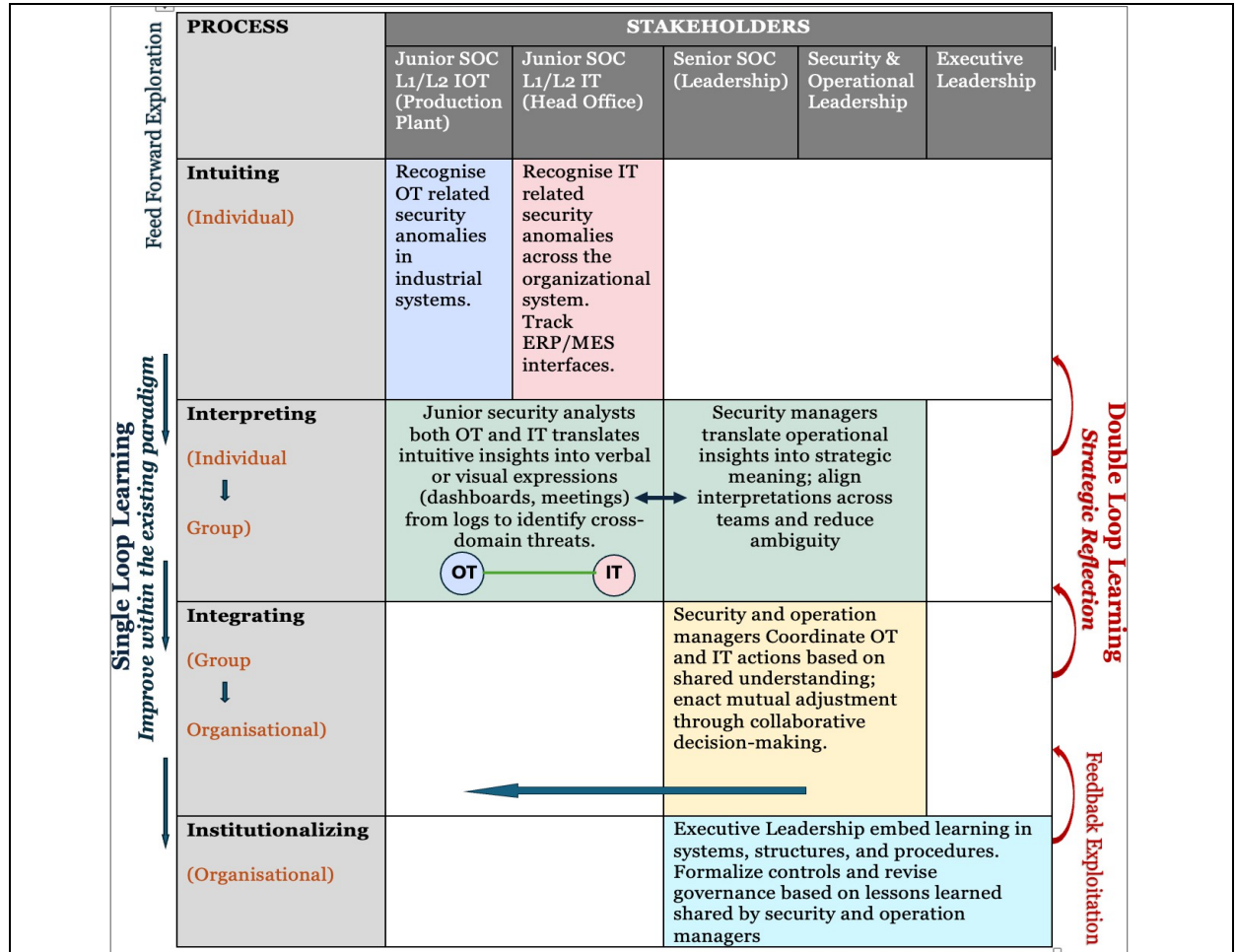

**Figure 1: An Advanced Dynamic Security Learning (DSL) Process Model**

### Intuiting

In the "Intuiting" phase, junior SOC L1/L2 experts (OT) identify security flaws in industrial control systems like sensor networks, SCADA, and PLCs in manufacturing plants (see **Figure 1**). Junior SOC L1/L2 teams at head office find odd ERP and MES logs as well as dubious user activity. This phase captures unconscious experiential insights and pattern recognition, initiating the learning loop. OT abnormalities in Industry 4.0 might have instantaneous and cascading physical impacts that call for quick discovery and careful analysis. The identification of certain security abnormalities marks the beginning of the Intuiting stage of the model. Here, the sense-making process begins with early discovery in smart industrial environments calls for considering corporate software, communication tools, and network traffic in IT as well as ICS, PLCs, SCADA, and other IoT devices in OT.

**Interpreting**

In interpreting, people and groups are linked. At this phase, the intuitive outputs of junior SOC teams are translated into communicable, understandable forms the organization can grasp. Junior SOC teams in IT and OT provide both verbal and graphic reports. Example includes real-time dashboards, log analytics, alert summaries, and risk indicators. Cross-functional meetings or SIEM systems help to distribute these artifacts. The strategic translation capacity of senior SOC and security management make this model unique. They provide a common understanding of the increasing risk to match the points of view of IT and OT. This approach removes uncertainty, corrects linguistic errors, and clarifies how deviations in one field could find their way to another.

**Integrating**

Through group decisions, the integrating phase operationalizes the shared knowledge from the previous phase. Design and implementation of mitigating techniques as well as transitions from group learning to organizational coordination constitute this phase. Important players include senior SOC team members, security managers, and operational leadership. Their shared knowledge drives their coordination of cross-domain incident response. This model stresses the synchronizing and reciprocal adjustment of IT and OT response strategies. For example, if a malware infestation in an OT subsystem connected to the corporate network, the IT team may isolate affected servers while OT oversees shutdown or manual operation. This phase consists of active single-loop learning as the firm improves configurations, monitoring thresholds, and procedural scripts without changing strategic assumptions.

**Institutionalizing**

Through the institutionalizing phase, lessons officially find their way into systems, policies, and procedures. At this phase, security and top leadership ensure that event lessons are recorded and incorporated into governance systems. Security directives, network architecture changes, access control changes, and SOP updates are finished during this phase. Reevaluating goals, assumptions, and even the organizational perception of risk, double-loop learning, also referred as strategic adaptation challenges and removes outdated assumptions. It may also point to a flaw in supplier screening, which would call for a modification to outside risk management protocols. Crossan et al. (1999) emphasize cycling learning back through strategic renewal. This phase also produces feedback (exploitation) for previous phases. Instituted lessons change the mental models, views of anomalies, and reactions to new hazards of junior SOC teams.

**Feedback (Double-Loop Learning) and Feedforward (Single-Loop Learning)**

The concept revolves around feedforward and feedback learning cycles. Though feedforward learning looks at learning from people to the organization, feedback learning employs institutional knowledge to impact individual actions, claims (Crossan et al., 1999). In the suggested advanced DSL model, single loop learning rules most of feedforward. This learning helps continuously improve monitoring systems, configurations and procedures. Conversely, especially in institutionalizing, the feedback loop shows double-loop learning. This process explicitly incorporates single- and double- loop learning. The number of connected IoT devices is expected to grow to 18.8 billion globally by the end of 2024, significantly expanding the attack surface and necessitating more sophisticated feedback and feedforward loops (Sinha, 2024). Industry 4.0 calls for scalable, practical, actionable learning models and this advanced DSL architecture supplies them.

*Implementation Approach*

The implementation plan of Advanced DSL model follows a three-phase approach spanning six months. Phase 1 (1-2 months): Assessment of current SOC capabilities of organization or establishment of independent SOC units for IT and OT environments if they do not exist in the organization. Phase 2 (3-4 months): Deployment through training staff on 4I framework principles and double-loop learning concepts. Implementation of cross-domain stakeholder post-incident reflection and using the real-time dashboards to support both IT and OT-related anomalies. Phase 3 (5-6 months): Institutionalize the learning outcomes into the governance systems and policies. Evaluate by measuring the organizational improvements in incident response. Success can be measured by the organization using the quantitative indicator such as 40% reduced incident response time and qualitative indicator such as assumption questioning (double-loop learning) evidence.

## Conclusion

This study proposes a scalable and consistent cybersecurity incident response model for Industry 4.0. The DSL model responds to rising cyber threats by using double-loop learning and the 4I organizational learning architecture to provide continuous, organization-wide learning and strategy renewal. Unlike Ahmad et al.'s original DSL model, the Advanced DSL model proposed here introduces a dual-domain SOC structure, integrates organizational reflection across OT and IT layers, and directly applies the 4I learning phases to operationalize learning across hierarchical levels in Industry 4.0. This extension provides a more robust and domain-specific architecture capable of addressing the cyber-physical convergence and real-time operational demands in smart industrial systems. The implications for cybersecurity practitioners are significant; in this model not only strengthens incident response effectiveness but also promotes long-term strategic alignment between cybersecurity operations and business resilience goals. Industry 4.0 organizations can use this model to bridge cross-functional gaps, institutionalize learning, and dynamically revise their governance based on reflective insights. To implement the Advanced DSL model effectively, organizations are recommended to establish independent SOC units for IT and OT environments to enhance early-stage anomaly detection and institutionalize structured post-incident reflection processes involving cross-domain stakeholder review panels.

### *Potential Research Contributions*

A major theoretical contribution of Advanced Dynamic Security Learning (DSL) fills in key gaps in cybersecurity incident response systems in Industry 4.0 environments. An important change is the focus on different but linked SOC duties for the OT and IT sectors. This captures the reality of Industry 4.0 cybersecurity, in which systems vary in behavior, risk, and response needs. The approach operationalizes feedforward (single-loop) and feedback (double-loop) learning to improve procedural responses, strategic thinking, and resilience, fitting for long-term adaptation in fast changing technology environments. This model focuses on the dynamic and decentralized nature of operational technology (OT) unlike the traditional incident response models that focus on IT-related anomalies with static digital assets (Lezzi et al., 2018). Practical industrial implementations will reveal that the model enhanced early threat detection, cross-functional collaboration, and strategic learning integration across IT and OT domains.

### *Research Limitations & Future Research Directions*

DSL is a great foundation, but it has some limitations. Empirically evaluating this conceptual model has not been done yet using case studies, simulations, or real event response testing. Applied to real-world organizational environments, practical complexity and unexpected implementation difficulties may arise. This model assumes organizations have dual SOC structures and capabilities for cross-functional learning which may not exist in all Industry 4.0 environments. For companies with scattered IT/OT systems or limited resources, using the approach might be challenging as many organizations do not provide cybersecurity training to all their employees (Hornetsecurity, 2024). The different incident response protocols and delayed communication due to the cultural gaps between the IT and OT teams might constrain the model's effectiveness.

Future studies will empirically validate the Advanced DSL model in industries operating in Industry 4.0 environments as this is the research-in-progress paper. The DSL model will be evaluated in Industry 4.0 settings through three planned case studies: one Italian smart industry and two Australian smart industries. The researcher's established networks in Italy and Australia, developed through three-year residency and ongoing PhD studies, provide access to incident response capabilities in smart industries and compare their different regulatory environments (EU GDPR vs. Australian cybersecurity frameworks). The main selection criteria of the industries are based on (1) they must have Industry 4.0 infrastructure (2) they must have established dual SOC teams and (3) a minimum 10 cybersecurity staff available. Data collection will be through the semi-structured interviews with SOC personnel and management, and participant observations during real incident response scenarios. Participants will be total 30: 10 cybersecurity professionals from the Italian organization and 20 from the Australian organization. Cybersecurity experts' focus group interviews are expected to corroborate the model's applicability by showing how well-organized learning loops enhanced incident response coordination and organizational flexibility, while assessing organizational boundary conditions and the barriers to implementation the Advanced DSL model. Case studies will reveal the approach may transform advanced anomaly detection into ongoing security enhancement. Research indicates that 65% of smart factories suffer ransomware attacks annually (Mahendru, 2024), these empirical validations make the DSL model a practical and dynamic cybersecurity resilience solution for rapidly growing smart industrial ecosystems.